\documentclass[prd,longbibliography,nofootinbib,preprintnumbers,twocolumn,showpacs,floatfix,superscriptaddress]{revtex4-1}

\usepackage[utf8]{inputenc}      
\usepackage{amsmath}
\usepackage{amssymb}
\usepackage{graphicx}
\usepackage{color}
\usepackage{hyperref}

\newcommand{\bitem}{\begin{itemize}}
\newcommand{\eitem}{\end{itemize}}
\newcommand{\bwt}{\begin{widetext}}
\newcommand{\ewt}{\end{widetext}}
\newcommand{\beq}{\begin{equation}}
\newcommand{\eeq}{\end{equation}}

\newcommand{\bdm}{\begin{displaymath}}
\newcommand{\edm}{\end{displaymath}}
\newcommand{\bea}{\begin{eqnarray}}
\newcommand{\eea}{\end{eqnarray}}

\begin{document}
\jot = 1.4ex         

\title{Gluon Mass Generation from Renormalons and Resurgence}

\author{Alessio Maiezza}
\affiliation{Dipartimento di Scienze Fisiche e Chimiche, Universit\`a degli Studi dell'Aquila, via Vetoio, I-67100, L'Aquila, Italy,}
\email{alessiomaiezza@gmail.com}

\author{Juan Carlos Vasquez}
\affiliation{Department of Physics $\&$ Astronomy, Amherst College, Amherst, MA 01002, USA}
\email{jvasquezcarmona@amherst.edu}

\begin{abstract}
We establish a link between the concepts of infrared renormalons, infrared fixed point, and dynamical nonperturbative mass generation of gluons in pure Yang-Mills theories. By utilizing recent results in the resurgent analysis of renormalons through non-linear ordinary differential equations, we develop a new description for the gluon propagator, thereby realizing the Schwinger mechanism. Specifically, this approach leads to a nonperturbative, dynamic mass generation for Yang-Mills gauge bosons in the deep infrared region, a phenomenon closely associated with color confinement. Furthermore, we present arguments about the limit of applicability of the Borel-Ecalle resummation of the renormalons by comparing it with the Kallen-Lehman representation of the gluon propagator.
\end{abstract}

\maketitle

\section{Introduction}

In Quantum Field Theory (QFT), a prevalent approach to bridging the gap between perturbative and non-perturbative phenomena involves examining the large-order behavior of the power series expansions in the coupling constant. This analysis, particularly focused on the points where renormalized perturbation theory becomes inadequate, may shed light on more complex features such as the generation of mass in Yang-Mills theories and the phenomenon of confinement. One of the hallmark indicators of perturbation theory reaching its limits is the emergence of singularities along the semi-positive axis in the Borel plane. These singularities, termed ``renormalons"~\cite{tHooft:1977xjm} and distinct from those associated with semi-classical instantons, are often associated with bubble diagrams -- for a detailed review about renormalons see Ref.~\cite{Beneke:1998ui}. More recent renormalon-based analyses are found in Refs.\cite{Abbas:2012fi,Maiezza:2018pkk,Cvetic:2019jmu,Correa:2019xvw,Boito:2021ulm,Loewe:2021ekj,Ayala:2021mwc,Loewe:2022aaw,Ayala:2022cxo,Caprini:2023tfa}.

In some cases, renormalons can also be obtained from the renormalization group equation (RGE)~\cite{Parisi:1978iq}. Along this line but exploiting resurgent properties of non-linear ordinary differential equations (ODEs)~\cite{Costin1995,costin1998,CostinBook}, a derivation from the RGE of the complete analytical structure of the renormalons has been proposed in Refs.~\cite{Bersini:2019axn}. Recent phenomenological applications in the context of QCD can be found in Refs.~\cite{Maiezza:2021mry,Maiezza:2021bed,Caprini:2023kpw}. The mathematical resurgence is a theory that enables to reconstruct the nonperturbative information from perturbation theory~\cite{Ecalle1993} -- for reviews, see Ref.~\cite{sauzin2007resurgent,Dorigoni:2014hea,Aniceto:2018bis}. Ideas from resurgence have been applied in QFT several times~\cite{Dunne:2013ada,Borinsky:2017hkb,Maiezza:2019dht,Clavier:2019sph,Borinsky:2020vae,Fujimori:2021oqg,Borinsky:2022knn,Laenen:2023hzu}.

In this paper, we explore the question of dynamical mass generation in Yang-Mills theories by applying techniques borrowed from resurgence theory, a novel method in this context. The approach centers on the study of ``renormalons," to bridge the gap between perturbative and nonperturbative QFT. We demonstrate that the resurgence of renormalons can realize the Schwinger mechanism~\cite{Schwinger:1962tn,Schwinger:1962tp} (non-perturbative dynamical mass generation) in pure gluon-dynamics. This mechanism is supported by lattice simulations~\cite{Bogolubsky:2007ud,Duarte:2016iko}.

Furthermore, Ref.~\cite{PhysRevLett.90.152001} argues that the generation of a dynamical gluon mass implies the presence of an IR fixed point~\footnote{For early discussions on dynamical mass generation, see Refs.~\cite{PhysRevD.26.1453}. Additional insights can be found in Refs.~\cite{Aguilar:2004sw,Aguilar:2004kt}, and for a comprehensive review, see Ref.~\cite{Aguilar:2015bud}.}. On the other hand, non-perturbative contributions coming from the Borel-Ecalle resummation of the renormalons can lead to a fixed point, avoiding Landau pole~\cite{Maiezza:2023mvb}.  Consequently, we are prompted to investigate the potential relationship between non-perturbative mass generation, renormalons, and the corresponding fixed point.

\section{Schwinger mechanism and renormalons}

Schwinger mechanism underlies a nonperturbative mass generation: the limit of vanishing four-momentum in the two-point function, of an originally massless field at the Lagrangian level, gives a nonzero constant value. This is an unequivocal sign of dynamical mass generation~\cite{Schwinger:1962tn}. We aim to realize this mechanism in pure Yang-Mills theory.

To this end, consider the RGE for the renormalized two-point green function of the massless gauge field.
Assume that there exists an IR fixed point at a non-zero coupling constant $\alpha(\mu_*)=\alpha_*$, such that $\beta(\alpha_*)=0$. Let us consider the two-point function in the Landau gauge (and Euclidean space). It is a straightforward exercise to show that the general solution to the RGE equation when $\beta(\alpha_*)=0$ is of the form~\cite{coleman_1985}
\begin{equation}\label{ColemanEq}
\Gamma^{(2)}_{\mu\nu}  = \left[\left(g_{\mu\nu} -\frac{p_{\mu}p_{\nu}}{p^2} \right)\, p^2\right] K(\alpha_*)\left(\frac{\mu_*^2}{p^2}\right)^{\gamma_*}\,,
\end{equation}
where we write the tree-level expression into the square parentheses, $K$ is an arbitrary function depending only on $\alpha_*$, and  $\gamma_*= \gamma(\alpha_*)$.  In particular, when $\gamma_*=1$ the factor $p^2$ in Eq.~\eqref{Green} cancels out, and the Green function takes the form (in the limit of small momentum $p$)
\begin{equation}
\lim_{p\rightarrow 0}\Gamma^{(2)}_{\mu\nu}=\left(g_{\mu\nu}-\frac{p_{\mu}p_{\nu}}{p^2}\right)K(\alpha_*)\mu_*^2\,,
\end{equation}
thus realizing the well-known Schwinger mechanism, consistent with the decoupling solution and gluon mass generation also discussed in Refs.~\cite{PhysRevD.26.1453,PhysRevD.40.3474,Aguilar:2004kt,Aguilar:2004sw,Frasca:2007uz,PhysRevLett.90.152001,Weber:2012vf}.

Therefore, within this setup, our goal is twofold: first, show the possibility of having an IR fixed point; second, see how to implement the condition $\gamma_*=1$.

\paragraph{Nonperturbative physics and renormalons.}
The generic tool to realize the above-proclaimed objectives is the Borel-Ecalle resummation of the IR renormalons, which are infinitely many singularities in the Borel space at
\begin{equation}\label{IR_renormalon_def}
z_{sing}= -\frac{2\,n}{\beta_1}\,,
\end{equation}
being $\beta_1$ the one-loop approximation of the beta function, and $n$ a positive integer.
Since $\beta_1$ is negative in the case of Yang-Mills theory, the singularities lie on the semi-positive axis in the Borel plane ($z_{sing}>0$), thus preventing the usual Borel-Laplace resummation. Mathematically, one can bypass this problem using a Borel-Ecalle resummation, as proposed in~\cite{Maiezza:2019dht,Bersini:2019axn}. As a result, the IR renormalons can be resummed in terms of a transseries containing a single unknown parameter. The latter remains undetermined due to the lack of a semiclassical limit of the renormalons. However, the generalized resummation implies a huge improvement concerning the infinitely many ambiguities one faces in the ordinary resummation. In addition, the \textit{a priori} unknown single-parameter-transseries can be adjusted by matching with phenomenology.

Specifically, our model relies on the following points:
\begin{itemize}

    \item We assume that the leading part of large order contributions to the anomalous dimension $\gamma(\alpha)$  can be estimated with the $n!$-growth coming from (IR) renormalons.

    \item We Borel-Ecalle resum these contributions, obtaining a nonperturbative approximation (in terms of a transseries).

\end{itemize}
We already commented on the second point. The first one is worth discussing. Indeed, an immediate question is about the meaning of renormalons in pure Yang-Mills models since all the skeleton diagram estimations center on fermion bubbles. The answer is that the singularities on the semi-positive Borel axis related to the one-loop beta function (namely IR renormalons, in the case of the Yang-Mills model) can also be seen directly from RGE, with a few plausible hypotheses~\cite{Bersini:2019axn}. As we shall summarize below, the crucial point is a nonlinear ODE, extracted from RGE and having precisely the properties of the renormalon in the Borel plane. More important, the ODE enables one to uniquely Borel-Ecalle resum the renormalons.

\section{Renormalization Scheme Invariance of the Renormalon's Resurgence Approach} \label{SecII}

In this section, we highlight the framework of the the resurgent approach to the renormalons~\cite{Bersini:2019axn,Maiezza:2023mvb}, necessary to deal with the Schwinger mechanism. We pay particular attention to the renormalization scheme independence of the method.

One basic point is to complete perturbation theory with a non-analytic function, $R(\alpha)$, whose Borel transform features singularities at integer multiples of $2/\beta_1$, being $\beta_1$ the one-loop coefficient of the beta function~\cite{Bersini:2019axn}.
In particular, we assume the function $R(\alpha)$ to appear in the finite part of the two-point correlator -- the finite part that is the one to be factorially divergent due to renormalons~\cite{tHooft:1977xjm}. It is worth emphasizing that a recent study~\cite{Balduf2023} on the Schwinger-Dyson equations confirms the appearance of $n!$ contributions in the finite part of the Green function, in agreement with the assumptions made in this paper.

Finally, $R$, identified with the Borel-Ecalle resummation of the renormalons can remove the Landau pole since it can be used to calculate non-perturbative fixed points~\cite{Maiezza:2023mvb}.

\paragraph{Renormalization group equation.}
Consider the 1-particle-irreducible two-point Green function (in Landau gauge and Euclidean space)
\begin{equation}\label{Green}
\Gamma^{(2)}_{\mu\nu} = \left[\left(g_{\mu\nu} -\frac{p_{\mu}p_{\nu}}{p^2} \right)\, p^2\right] \, \Pi(p^2,\mu^2) \,,
\end{equation}
where $p$ is the four-momentum.

Defining $L:=\log(\mu_0^2/\mu^2)$, with $p^2:=\mu_0^2$, the vacuum polarization function $\Pi$
in Eq.~\eqref{Green} satisfies the RGE
\begin{equation}\label{CS}
\left[ -2 \frac{\partial}{\partial L} + \beta(\alpha) \frac{\partial}{\partial_\alpha} - 2 \gamma(\alpha) \right]  \, \Pi(L) = 0 \,,
\end{equation}
with $\beta(\alpha) = \mu \frac{d\alpha}{d\mu}$.

One can write the vacuum polarization function in
the scale expansion form
\begin{equation}\label{PT_ren}
\Pi(L)=  \sum_{k=0}^{\infty} \pi_k(\alpha) L^k \,,
\end{equation}
showing that RGE corresponds to an infinite system of ODEs~\cite{KreimerYeats2006,KreimerYeats2008,vanBaalen:2008tc,Yeats2008,Kreimer2008,vanBaalen:2009hu,Klaczynski:2013fca}.

By the renormalization conditions, the finite part of the above Green function can be written as $\pi_0=1$ at any order in perturbation theory. If one implements this at order $N$, one makes an error of the order of  $\alpha^{N+1}$.  However, due to the $n!$ divergence of perturbation theory, the latter procedure does not hold when $N\rightarrow\infty$.
Therefore, the condition $\pi_0=1$ is not well-defined at the non-perturbative level, and one expects that $\pi_0$ must be completed with a non-perturbative function, $R(\alpha)$, in line with the concept of renormalon (as recalled at the beginning of this section). Therefore our \emph{ansatz} is~\cite{Bersini:2019axn}
\begin{equation}\label{introduce_R}
\pi_0(\alpha)=1+R(\alpha) \,.
\end{equation}
Plugging Eq.~\eqref{PT_ren} into Eq.~\eqref{CS} and equating to zero the coefficients of the term $L^n$, one gets the following infinite set of differential equations
\begin{equation} \label{pi-resursion}
2 \gamma (\alpha ) \pi_k(\alpha)  +2 (k+1) \, \pi_{k+1}(\alpha ) = \beta (\alpha ) \pi'_k(\alpha) \,,
\end{equation}
with $k=0,1,2,...$.  Consider now the equation when $k=0$, namely
\begin{equation}\label{pi0-equation}
2 \gamma (\alpha ) \pi_0(\alpha)  +2  \, \pi _{1}(\alpha ) = \beta (\alpha ) \pi'_0(\alpha) \,,
\end{equation}
and plugging Eq.~\eqref{introduce_R}  into Eq.~\eqref{pi0-equation} one gets
\begin{equation}\label{ODE}
2 \gamma (\alpha ) (1+R(\alpha))  +2  \, \pi _{1}(\alpha ) =  \beta (\alpha ) R'(\alpha) \,,
\end{equation}
Both $\gamma$ and $\beta$ must also depend on $R$: by solving RGE, the knowledge of $\gamma$ and $\beta$ would completely determine the two-point correlator, which depends on $R$. When $R\rightarrow 0$, the anomalous dimension $\gamma(\alpha)=-\pi_1(\alpha):=\gamma_{pert}$, as it should be and consistently with Eq.~\eqref{ODE}.

Thus, in general, one can formally write
\begin{align}
\gamma(\alpha) &= \gamma_{pert}(\alpha) +f_{\gamma}(R)\,, \label{betaNP2}\\
\beta(\alpha) &= \beta_{pert}(\alpha) +f_{\beta}(R)\,,  \label{betaNP1}
\end{align}
such that both the \textit{a priori} unknown functions $f_{\gamma},f_{\beta}\rightarrow 0$ when $R\rightarrow 0$, and where $ \gamma_{pert}$ and $ \beta_{pert}$ denote the perturbative expressions for the $\gamma$ and the $\beta$ functions:
\begin{align}\label{betagammapert}
\gamma_{pert}(\alpha) & = \gamma_1 \alpha +\gamma_2 \alpha^2 +...\,,\\
\beta_{pert}(\alpha) & = \beta_1\alpha^2 +\beta_2\alpha^3 + ...\,. \label{betagammapert2}
\end{align}
In what follows, we shall assume that the function $R$ enters linearly in $\gamma$, and we label this as the \emph{minimal setup}.
Thus we have~\footnote{For convenience, we have made a change of sign in the linear term concerning Ref.~\cite{Maiezza:2023mvb}.}
\begin{align}\label{gammanonpert}
\gamma(\alpha) &= \gamma_{pert}(\alpha) -q R\,, \\
\beta(\alpha) &= \beta_{pert}(\alpha) -b\,\alpha R\,. \label{betanonpert}
\end{align}
By matching with the one-loop Landau pole structure, UV and IR respectively, the parameter $q$ is +1 for asymptotically free models (and -1 for non-asymptotically free models)~\cite{Bersini:2019axn,Maiezza:2023mvb}. Since we discuss Yang-Mills models in this work, we set $q=1$ from here on.

A comment on Eq.~\eqref{betanonpert} is in order. The appearance of $R$, so the renormalons, in the beta function may result unfamiliar to the Reader. Nevertheless, the absence of renormalons into $\beta$ is just a feature of the $\overline{MS}$-scheme, while they are present in on-shell-scheme~\cite{Broadhurst:1992si} -- see also the discussion in Ref.~\cite{Bersini:2019axn}.

\paragraph{A non-linear ODE.}
Upon substitution of the leading term in Eq~\eqref{betagammapert2} into Eq.~\eqref{ODE} and changing variable $\alpha=1/x$,  one gets
\begin{align}\label{ODE-one}
 R'(x) &= F(1/x,R(x))=\frac{\frac{2}{\beta_1}  -\frac{2}{\beta_1}\gamma(x) }{1-\frac{b}{\beta_1}x\, R(x)}R(x) \,.
\end{align}
Next, by replacing $\gamma(x)$ using Eq.~\eqref{gammanonpert}, we bring the ODE above to its normal form -- as presented in Sec. (5.4) of Ref.~\cite{CostinBook}.
To this end, we perform first the change of variables
\begin{equation}
R(x) = \frac{U(x)}{x} = \alpha U(\alpha)\,
\end{equation}
and Eq.~\eqref{ODE-one} becomes
\begin{align}\label{ODE-final}
U(x)'  = &  -Q\,U(x) + A\frac{U(x)}{x}+\mathcal{O}(1/x^2,U(x)^2)\,,  \\
&\text{where,}\, \nonumber \\
& Q = -\frac{2}{\beta_1}\,,\text{ and } A= \frac{\beta_1-2 \gamma_1}{\beta_1}\,.\label{QA}
\end{align}
Finally, one shifts $\bar{U}(x) = U(x)+\mathcal{O}(1/x^{N+1})$, being $N$ large enough to have a formally small shift. This yields
\begin{align}\label{ODE-final-final}
\bar{U}(x)'  = &  -Q\,\bar{U}(x) + A\frac{\bar{U}(x)}{x}+ \mathcal{O}(1/x^{N+1})\nonumber \\ &+\mathcal{O}(1/x^2,\bar{U}(x)^2)  \,.
\end{align}
Considering higher-loop corrections in Eq.\eqref{ODE-one} expansion (e.g. $\beta_2$) only modifies the term $A$ as follows
\begin{equation}\label{new_A}
A= \frac{\beta_1^2-2\beta_1\gamma_1+2 \beta_2}{\beta_1^2} \,,
\end{equation}
and the meaning of $A$ is discussed in more detail in the next paragraph.

\paragraph{Analytic properties of the Borel transform.}
The previous (kind of) equation is important for its properties of the Borel transform $B(\bar{U}(x))$ of $\bar{U}(x)$: the non-linearity in $\bar{U}$, via self-convolution, gives infinitely many singularities at (see e.g. Ref.~\cite{CostinBook})
\begin{equation}
z_{sing}= n\, Q= -\frac{2\,n}{\beta_1} \,\,\,\,\,\,\,\, (n\,\, \text{is a positive integer})\,.
\end{equation}
The same holds for $B(R(x))$. Notice that we do not explicitly write the non-linearity in Eq.~\eqref{ODE-final-final} since the specific form is irrelevant, but their presence is pivotal.

The crucial point is that these singularities match with the one in Eq.~\eqref{IR_renormalon_def}. Therefore, $\bar{U}(x)$ can be identified with the resummation of IR renormalons.

The  coefficient $A$ controls the type of poles in the Borel transform of $\bar{U}(x)$, namely whether the singularities in the Borel transform are simple, quadratic poles or, in general, branch points:
\begin{equation}\label{typepole}
B(\bar{U}(x)) \propto \frac{1}{\left(z-Q\right)^{1+A}}\,.
\end{equation}
As shown in Eq.\eqref{new_A}, higher loop corrections only affect the kind of singularities in the Borel transform of the Green function.

\paragraph{Resurgence of the renormalization group equation.}
Let us make manifest the resurgent properties of the general solution of Eq.~\eqref{ODE-final-final}, which is of the form~\cite{Costin1995}~\footnote{The transseries solution can also be re-obtained in the Ecalle's language of alien calculus~\cite{Maiezza:2023mvb}.}
\begin{equation} \label{transseries-sol}
\bar{U}(x) = \sum_{n=0}^{\infty} C^n \bar{U}_n(x)e^{-\frac{2nx}{\beta_1}}\,,
\end{equation}
which is a single-parameter transseries. Note that since Eq.~\eqref{ODE-final-final} is a first-order ODE, there is only one arbitrary parameter, $C$.

The function $\bar{U}_0(x)$ is the result one obtains from perturbation theory, and all the other $\bar{U}_n(x)$ can be recursively calculated as
\begin{equation}\label{costin_mult}
\bar{U}_n(x) = e^{n  x}\left(\delta_{0}\bar{U}_0(x) -\sum_{j=1}^{n-1} e^{-j x} \bar{U}_j(x)	\right) \,\,\,, n\geq1 \,,
\end{equation}
where $\delta_0$ is the discontinuity calculated along the positive real axis in the Borel complex plane.  This expression shows the resurgence within the ODE formalism.

Remarkably, as demonstrated in Eq.~\eqref{QA}, the non-perturbative constants $Q$ and $A$ depend solely on the coefficients $\gamma_1$, $\beta_1$ and $\beta_2$. These coefficients are notable for their independence from the renormalization scheme. The higher-order coefficients $\beta_3, ...$ and $\gamma_2, ...$, however, come into play at higher orders, specifically in terms of $\bar{U}(x)^2$ or $1/x^2$ and higher. Consequently, these coefficients do not alter the position or the type of singularity in the Borel transform of the Green functions. This characteristic further confirms the renormalization scheme independence of $Q$ and $A$ in Eq.~\eqref{ODE-final-final} and the transseries solution in Eq.~\eqref{transseries-sol}. Finally, note that invariance under changes in the renormalization scale $\mu$ is, by construction,  automatically guaranteed for any function satisfying the RGE.

\paragraph{A Non-perturbative Fixed Point.}
The Eqs.~\eqref{betaNP1} and~\eqref{betanonpert} show the possibility of having a non-perturbative fixed point due to $R$.

Notice that one can turn around the argument: one can introduce a non-perturbative function $R$ to eliminate the presence of the (perturbative) Landau pole, thus for the sake of consistency of the theory\cite{Maiezza:2023mvb}.

\section{A model for dynamical mass generation in pure Yang-Mills theory} \label{SECIII}

In this section, profiting from the concepts developed above, we show how the Schwinger mechanism~\cite{Schwinger:1962tn,Schwinger:1962tp} can be realized through a combination of renormalon-inspired considerations and resurgence analysis.

We shall see that the resurgence of renormalons in the minimal setup (Eqs.~\eqref{gammanonpert}  and~\eqref{betanonpert}) leads, for the anomalous dimension at the fixed point, to satisfy $\gamma_*=1$.

\paragraph{Resurgent renormalons and nonperturbative mass.}
We now turn our attention to the prediction that one gets for the solution shown in Eq.~\eqref{ColemanEq} within the resurgences's formalism for renormalons. To this end, note that when $\beta(\alpha^*)=0$, the set of differential equations in Eq.~\eqref{pi-resursion} reduces to an infinite set of algebraic equations that can be solved analytically. In particular
\begin{equation}\label{ODE-eqs}
 \gamma (\alpha^*  ) \pi_k(\alpha^* ) +  (k+1) \, \pi _{k+1}(\alpha^*  ) = 0 \,,
\end{equation}
where the solution to the above set of recursive relations is
\begin{equation}\label{eq-recurrence}
    \pi_k^* =\pi_0^*  \frac{(-1)^k(\gamma^* )^k}{k!}\,.
\end{equation}
Replacing in Eq.~\eqref{PT_ren}, one gets
\begin{equation}\label{ourEQ}
\Pi(L)= \pi_0^* e^{-\gamma^* L} =\pi_0^*\left(\frac{\mu_*^2}{\mu_0^2}\right)^{\gamma^*}\,.
\end{equation}
By comparing Eq.~\eqref{ColemanEq} with Eq.~\eqref{ourEQ} and choosing $\mu_0^2=p^2$, one finds
\begin{equation}
K(\alpha^*) = \pi_0^*.
\end{equation}
Note, the function $K(\alpha_*)$ depends non-analytically on the coupling $\alpha$, consistently with a dynamically generated mass~\cite{Balian:1976vq}.
Thus far, the treatment is general, as well as the solution to the RGE. In what follows, we are going to show that the \emph{minimal setup} shown in Eqs.~\eqref{gammanonpert} and \eqref{betanonpert} predicts $\gamma^*=1$.

We now turn our attention to the case of renormalons (the function $R$) and see what are the consequences we get from it.
Eqs.~\eqref{introduce_R} and~\eqref{eq-recurrence} for $k=0$ implies
\begin{equation}\label{eq-g-fixed-point}
\pi_0^* = 1+R^* =- \frac{\pi_1^*}{\gamma^*}\,.
\end{equation}
Once again, we adhere to the shorthand notation $R(\alpha^*)=R^*$. The relation shown in Eq.~\eqref{gammanonpert} is valid in general and hence also valid at the fixed point,  and (recall $q=-1$ for Yang-Mills models)
\begin{equation}\label{eq-g1-g-R-2}
    \gamma^* = -\left(\pi_1^*+ R^*\right)\,.
\end{equation}
  Finally both Eqs.~\eqref{eq-g-fixed-point} and \eqref{eq-g1-g-R-2} are satisfied simultaneously when
\begin{equation}\label{eq-q-fixing-2}
     \gamma^* = 1 \,.
\end{equation}
Therefore, after setting $\mu_0^2=p^2$,  the minimal setup gives,  at the leading order, a gauge boson mass (decoupling solution for confinement), and  one finds
\begin{align}\label{GammaatFP}
    \lim_{p\rightarrow 0}\Gamma^{(2)}_{\mu\nu} &=\left(g_{\mu\nu}-\frac{p_{\mu}p_{\nu}}{p^2}\right)(1+R^*)\mu_*^2 \nonumber \\
    &=\left(g_{\mu\nu}-\frac{p_{\mu}p_{\nu}}{p^2}\right)M^2_{g}\,.
\end{align}
where $M_g^2= \left(1+R^*\right)\mu_*^2$. This result is consistent with the one found using QCD on the lattice estimates~\cite{Bogolubsky:2007ud,Duarte:2016iko} -- for a concise review see Ref.~\cite{Aguilar:2015bud}.
Both the existence of an IR fixed point and the resurgent approach to the RGE are crucial in deriving this result.

Finally, note that there exists in the literature an alternative way of defining the coupling constant down to $p^2\rightarrow 0$. More specifically, $\alpha_s$ is defined as a product of (Landau gauge) dressing functions, proportional to the gluon dressing function~\cite{Bogolubsky:2009dc,Ayala:2017tco}. In this alternative non-perturbative approach there is no dynamical gluon mass generation, but rather a specific non-perturbative behavior of the running coupling in the deep infrared regime.

 \begin{figure}[t]
 \centerline{
\includegraphics[width=.95\columnwidth]{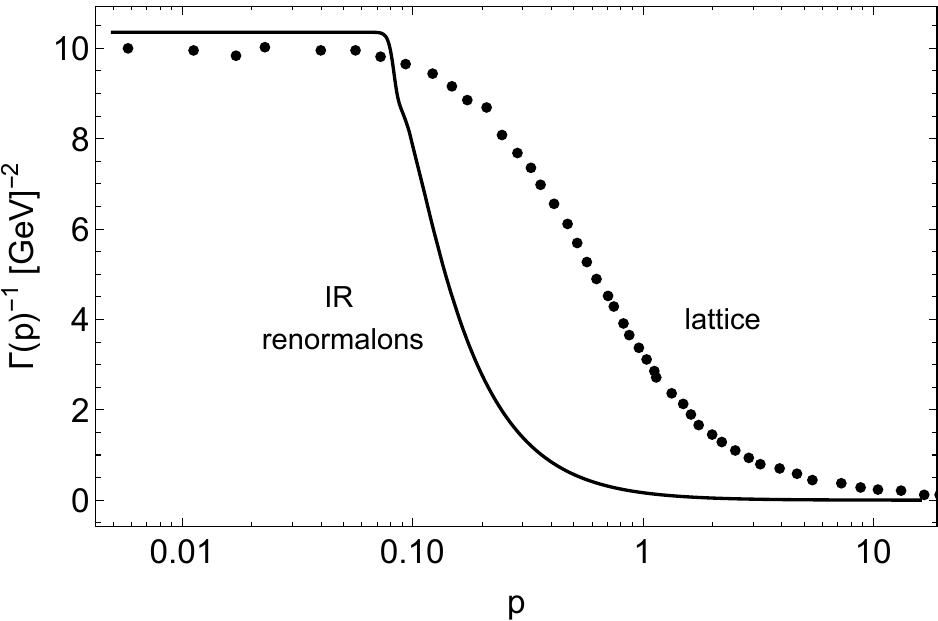}
}
\caption{The solid line represents the gluon propagator in Eq.~\eqref{gamma_plot}; the dotted line represents the lattice values taken from Ref.~\cite{Bogolubsky:2009dc}. }
\label{F}
\end{figure}

\section{An illustrative example for the gluon dynamical mass generation}\label{illustrative_model}

Knowing the form of the two-point function at the fixed point in Eq.~\eqref{GammaatFP}, one can calculate the running of the gauge propagator at any energy via Eq.~\eqref{CS}. We shall focus on the $SU(3)$ color gauge model.

We aim to provide an illustrative model of the IR gluon propagator and confront it with known lattice results. To this end, it is sufficient to implement Eq.~\eqref{gammanonpert} with the one-loop expressions for the $\beta$ and $\gamma$ functions, plus the nonperturbative contributions from the IR renormalons. Moreover, due to the IR renormalons, we assume the Borel structure just as an infinite sum of simple poles with alternate signs (as in Ref.~\cite{Maiezza:2019dht}), namely, we set $A=0$ in Eq.~\eqref{typepole}. Conceptually, the latter removes the term $\propto A$ in Eq.~\eqref{ODE-final}. The hypothesis of simple poles, instead of algebraic branch points, shall not drastically affect the result, as far as $\alpha^*$ is sufficiently small~\cite{Maiezza:2023mvb}.

We show our result in Fig.~\ref{F}. The solid line represents the propagator resulting from our model (from Eq.~\eqref{Green} and $p^2=\mu_0^2$):
\begin{equation}\label{gamma_plot}
\Gamma(p):= p^2 \Pi \left(\log \left(\frac{p^2}{\mu_*^2}\right)\right) \,,
\end{equation}
for $p^2\geq \mu_*^2$, and it has the constant value $M_g^2$ from Eq.~\eqref{GammaatFP}) below the fixed point, $p^2< \mu_*^2$.

We compare with the dots corresponding to lattice points taken from Ref.~\cite{Bogolubsky:2009dc}. Notice that to calculate the propagator in Fig.~\ref{F}, we need only two \textit{a priori} arbitrary constants, $C$ and $b$, the latter being contained in Eq.~\eqref{gammanonpert} and $C$ is the transseries parameter in Eq.~\eqref{transseries-sol}. We fix them choosing $\alpha^* \simeq 0.95$, corresponding to two conditions -- so fixing two parameters -- which are $\beta(\alpha^*)=0$ and Eq.~\eqref{eq-q-fixing-2}. In particular, we obtain $C\approx 0.8$ and $b\approx 48$. The choice of the $\alpha^*$ value aims to approximate the IR plateau of the lattice points (Fig.~\ref{F}).

In summary, the result in Fig.~\ref{F} qualitatively captures the behavior predicted by lattice simulations, despite all simplifying assumptions that do not enable us to quantitatively agree with the simulated propagator. Improvements, however, are possible: first, one should consider higher loop contributions for $\beta_{pert}$ and $\gamma_{pert}$; second, one can consider a more general pole structure of the Borel transform of the Green function in Eq.~\eqref{typepole}. This in-depth analysis is beyond the scope of the present work. Furthermore, even if one made the aforementioned improvements, it is far from clear whether one can get a quantitative agreement with the lattice results at all since we are neglecting all the other non-perturbative contributions -- e.g. instantons. Dealing with these together with renormalons is an open problem since one would have two superimposed Stokes lines (on the semi-positive axis), leading to the phenomenon of ``resonance" -- see Ref.~\cite{CostinBook}. Nevertheless, since the renormalons are the leading singularities in the Borel space -- the first renormalon is closer to the origin than the first instanton -- one may expect that the nonperturbative proposal is conceptually meaningful~\cite{Parisi:1978iq}.

\section{Superexponential Behavior of the Laplace Transform and Multiparticle Effects}\label{Sec:KL}

In the previous sections, we have implemented the Borel-Ecalle resummation of the renormalons, which assumes that the Laplace integral exists, namely
\begin{equation}\label{LapTras}
\Pi(\alpha) = \int_0^{\infty}\, dz\, e^{-\frac{z}{\alpha}} B[\Pi](z) <\infty\,,
\end{equation}
where $B$ denotes the Borel transform. However, 't Hooft argued that the Laplace integral is divergent due to the super-exponential behavior of the Borel transform~\cite{tHooft:1977xjm}~\footnote{In this context, accelero-summation proposed in Ref.~\cite{Ecalle1993} becomes of interest, and an application in QFT is in Ref.~\cite{Bellon:2018lwy}.}.

The argument on the divergence of the Laplace integral is rooted in the Kallen-Lehmann (KL) representation of the Green function and the \emph{multiparticle state singularities}. Specifically, the KL analytic structure for complex momenta implies singularities of the Green function at $\Pi=\Pi(\alpha_{sing})$, with
\begin{equation}\label{thooft_sing}
\frac{1}{\left[\alpha(k^2)\right]_{sing}} =   r +  \frac{\beta_1}{2}(2n+1) \pi i  \,.
\end{equation}
where $r$ is a real and \emph{arbitrarily large} constant.

The singularities at $\Pi(\alpha_{sing})$ implies that $\Pi(\alpha)$  has an accumulation point of singularities at $\alpha\rightarrow 0$, and then the Borel transform grows faster than any exponential~\cite{tHooft:1977xjm}.

Conversely, renormalons -- or the function $R(\alpha)$ -- are related to a subexponential behavior on the Borel transform and, in the entire treatment we neglect any effect that would make the Laplace integral in Eq.~\eqref{LapTras} divergent. This assumption is justified if one is not interested in processes involving multiparticle states. Therefore, we argue that the renormalon-based model for the Schwinger mechanism is reliable.

\section{Outlook}\label{Outlook}

After showing that the resurgent approach to renormalons gives results that remain invariant across different renormalization schemes, we have provided evidence that the \emph{minimal setup} can implement the Schwinger mechanism in the context of Yang-Mills theories, generating a dynamical gluon mass.

In the literature, this is achieved within various methods, among others with lattice-based estimates, however, no connection between the Schwinger mechanism and perturbation theory was known. In this work, we attempt to fill this gap using  Borel-Ecalle resummation applied to IR renormalons, objects that edge the limitations of perturbation theory.

Renormalons, as noted long ago in Ref.~\cite{Aglietti:1995tg}, turn out to properly ``trace" genuine nonperturbative effects. To push forward the use of renormalons, the specific key ingredient has been the resurgence of ODEs, enabling us to attain an IR fixed point alongside a dynamical mass generation that, in turn, has been argued to be a necessary signal of color confinement~\cite{Chaichian:2006bn}.

Finally, checking our results against known features coming from the Kallen-Lehmann representation of the gluon propagator, we argue in favor of the robustness of renormalon-based proposal for dynamical mass generation, as far as multiparticle states do not play any relevant role.

\section*{Acknowledgements}
AM thanks Fabrizio Nesti for discussions. JCV thanks the Amherst College physics and Astronomy department where this research was carried out.

\bibliographystyle{utphysmod}
\bibliography{biblio}

\end{document}